\documentstyle[11pt]{article}
\title{\bf Nucleus-nucleus interaction in the perturbative QCD}
\author{M.A.Braun \thanks{ 
Dep. High-Energy physics,
S.Petersburg University, 198504 S.Petersburg, Russia}\\
North Carolina Central University, Durham, NC, USA}
\date{}
\def\beq{\begin{equation}}
\def\eeq{\end{equation}}

\def\phid{\phi^{\dagger}}

\begin{document}
\maketitle
\medskip
\vspace{1 cm}

{\bf Abstract}

Nucleus-nucleus interaction is studied in the framework of the perturbative
QCD with $N_c\to\infty$ and a fixed coupling constant. The pomeron tree diagrams
are summed by an effective field theory. The classical field eqations are solved
by iteration procedure, which is found convergent in a restricted domain of not
too high energies and atomic numbers. The found gluon distributions do not scale,
have their maxima close to 2 GeV independent of rapidity and fall towards
the central rapidity region. The cross-sections slowly grow with energy due
to the contribution from peripheral collisions, where evolution remains
linear.
Simple variational estimates at higer rapidities confirm this tendency.
\vspace{1 cm}

\section{Introduction}
As discussed in ~\cite{bra1} in the perturbative QCD with a large number of 
colours $N_c$ and a fixed coupling
constant high-energy nucleus-nucleus interaction  is described by the
exchange of 
an arbitrary number of BFKL pomerons, which interact between themselves via
the three-pomeron coupling. Corrections due to interactions not reducible
to pomeron exchange but rather to gluonic exchange are of the order
$1/N_c^2$.
The resulting pomeronic diagrams can be classified according to the number of pomeronic
loops. Each pomeronic loop gives an additional factor $1/N_c^2$. So in the high-colour
limit only tree diagrams survive. In the case of the scattering on the nucleus of a
very smal probe (e.g a highly virtual photon) this leaves only pomeronic fan diagrams,
which can easily be summed to lead to the non-linear BFKL evolution equation
~\cite{bal,kov,bra2}.  
This equation, although not soluble analytically, can be comparatively 
easily solved 
by numerical methods (e.g.  ~\cite{bra2,lev,arm}).

For nucleus-nucleus scattering the situation complicates enormously. 
The basic complication comes from the fact that now the pomerons not only
split into two but also merge from two to one.
The tree diagrams now do not reduce to fans but involve other structures,
like shown in Fig. 1. Still, using methods of the effective
 non-local field theory, one
can sum all these diagrams, reducing the problem to the solution of a pair
of non-linear field equations in the 
rapidity-transverse-momentum space ~\cite{bra1}.
Unfortunaltely, contrary to the non-linear BFKL equation, these are not
 evolution equations but rather correspond to a system
of  full-fledged non-linear integral equations, which are very difficult to
solve.

In this paper we make a first attempt at a solution of these equations
and try to gain some insight into the physical picture of the
nucleus-nucleus interaction in this approach.
We use two different methods of  solution. First we try to find the
solution by iterative methods starting from the fan digrams only.
Unfortunately our results show that this method is convergent in a 
very restricted region of not too high rapidities and not too large
nuclei. Beyond this region the iterations do not converge, indicating some
sort of qualitative change in the  form of the solution (a phase
transition?). In relation to this it is worth remembering that the
primitive Glauber approximation formula for the nucleus-nucleus scattering
in the tree approximation shows a similar singularity as the nuclei become 
heavy enough (at $A=64$ for nuclei with a constant profile function within
their transverse areas)~\cite{pak}. To move beyond the mentioned limits we 
tried to
use a direct variational method to find the stationary point of the
effective 
action, choosing the simplest form of the trial fields. Comparison with
the exact solution where it can be found by oterations shows that the
precision of our variational results is not high (of the order $\sim$30\%).
Still it gives a possibility to see the qualitative behaviour of the
solutions at very high rapidities.

Our results show that in the nucleus-nucleus collisions the rise of the effective 
number of gluons becomes still more suppressed than in the non-linear BFKL equation
case. In fact the variational estimates indicate that it may even go down with
the growth of the rapidity. However this does not clearly reflect itself on the
final nucleus-nucleus cross-sections, which continue to slowly rise due to
the contribution from the peripheral parts of the nuclei, where, due to the
the small nuclear density, the evolution remains practically linear.

In general the effect of the pomeron interaction on the nuclear cross-section 
is not very impressive. This is a consequence of the fact that the 
nucleus-nucleus
amplitude gets automatically unitarized due to cancellations between
contributions of different disconnected parts. A much greater change can be
seen in the contribution of each such part (the eikonal function), which
at high rapidity becomes many orders of magnitude smaller than in the
pure linear BFKL evolution case. 

The paper is organized as follows. In Section 2. we remind our basic formalism
to treat the nucleus-nucleus collisions in the perturbative QCD approach,
which reduces the problem to searching for a stationary point for the
action of a certain
non-linear and non-local field theory 
In Section 3 we outline our methods to find this stationary and to solve 
the corresponding variational field equations. Section 4 presents our numerical results
which are discussed in Section 5.

\section{AB-cross-sections and effective field theory}

At fixed overall impact parameter $b$ and (high) rapidity $Y$ the 
 nucleus-A-nucleus-B total 
cross-section is given by
\beq \sigma(Y,b)=2\left(1-e^{-T(Y,b)}\right).
\eeq
Here the eikonal function $T$ is a contribution from the connected part and is
an integral over two impact parameters $b_A$ and $b_B$ of the collision point relative
to the centers of the nuclei A and B:
\beq
T(Y,b)=\int d^2b_Ad^2b_B\delta^2(b-b_A+b_B)T(Y,b_A,b_B).
\eeq

In the perturbative QCD, in the large $N_c$ limit, the eikonal function
is given by
a sum of all connected tree diagrams constructed of BFKL pomerons, which
interact
between themselves via the triple pomeron vertex (with a minus sign). It
can be shown that this sum
is generated by an effective field theory of two fields $\phi(y,q)$ and $\phi^{\dagger}(y,q)$
depending on rapidity $y$ and transverse momentum $q$ with an appropriately
chosen action $S$ ~\cite{bra1}. The action  consists of
a free part $S_0$, interaction part $S_I$ and external part $S_E$. The free part is 
given by
\beq
S_0=2\langle\phid|K\left(\frac{\partial}{\partial y}+H\right)|\phi\rangle
\eeq
where $H$ is the forward BFKL Hamiltonian for the so-called semi-amputated
amplitudes ~\cite{lip}
and $K$ is a differential operator in $q$ commuting with $H$
\beq 
K=\nabla_q^2q^4\nabla_q^2.
\eeq 
Symbol $\langle...\rangle$ means integrating over $y$ and $q$ with weight $1/(2\pi)^2$ 
Action $S_0$ generates propagators which are BFKL Green function with operators $K^{-1}$
attached at their ends.
The interaction part of the action describes splitting and merging of pomerons:
\beq
S_I=\frac{4\alpha_s^2N_c}{\pi}
\langle\Big({\phid}^2K\phi+\phi^2K\phid\Big)\rangle.
\eeq
The coefficient in this term depends on the normalization of the fields.
Finally the external action is
\beq
S_E=-\langle\Big(w_A\phi+w_B\phid)\Big)\rangle,
\eeq
where $w_{A,B}$ describe the interaction of the pomerons with the projectile and target.
If the colour distribution in the target is given by
\beq
\rho_A(r)=g^2AT_A(b_A)\rho(r),
\eeq
where $\rho(r)$ is the colour distribution in the nucleon and $T_A$ is the target nucleus
profile function, then
\beq
w_A(y,q)=\delta(y)\int d^2rr^2\rho_A(r)\equiv\delta(y)\hat{w}_A(q),
\eeq
the $\delta$ function indicating that the target is taken to be at zero rapidity.
Function $w_B(y,q)$ is given by a similar formula with $\delta(y)$ substituted
by $\delta(y-Y)$ wher $Y$ is the rapidity of the projectile. 

The classical equations of motion which follow, multiplied  by $(1/2)K^{-1}$ 
from the left, are
\beq
\left(\frac{\partial}{\partial y}+H\right)\phi+\frac{2\alpha_s^2N_c}{\pi}
\Big(\phi^2+2K^{-1}\phid K\phi\Big)=\frac{1}{2}K^{-1}w_A
\eeq
and
\beq
\left(-\frac{\partial}{\partial y}+H\right)\phid+\frac{2\alpha_s^2N_c}{\pi}
\Big({\phid}^2+2K^{-1}\phi  K\phid\Big)=\frac{1}{2}K^{-1}w_B.
\eeq
From the $\delta$-like dependence on $y$ of the external sources it follows
that the equations can be taken homogeneous in the interval $0<y<Y$,
action of the external sources substituted by the boundary conditions
\beq
\phi(0,q)=\frac{1}{2}K^{-1}\hat{w}_A(q),\ \  
\phid(Y,q)=\frac{1}{2}K^{-1}\hat{w}_B(q).
\eeq

The eikonal function $T(Y,b_A,b_B)$ is just the action $S$ calculated with the
solutions of Eqs. (9)and (10), $\phi_{cl}$ and $\phid_{cl}$:
\beq
T(Y,b_A,b_B)=-S\{\phi_{cl},\phid_{cl}\}.
\eeq
Using the equation of motions one can somewhat simplify the expression for $S$.
Indeed multiplying the first equation by $2K\phid$, the second one by $2K\phi$,
integrating both over $y$ and $q$ and summing the results one obtains a relation
\beq
2S_0+3S_I+S_E=0,
\eeq
which is valid for the classical action, that is, calculated with the
solutions of Eqs. (9) and (10). Using this relation we can exclude, say, $S_0$ 
from (12) to find
\beq
T(Y,b_A,b_B)=\frac{1}{2}\Big(S_I\{\phi_{cl},\phid_{cl}\}
-S_E\{\phi_{cl},\phid_{cl}\}\Big).
\eeq
The dependence on $b_A$ and $b_B$ comes from the boundary conditions (11).

\section{Methods of solution}
\subsection{Final formulas for calculation}
To solve Eqs. (9 and (10) we first rescale the rapidity and fields to pass to variables known
from studying fan diagrams ~\cite{kov,bra2}:
\beq
y\to y/\bar{\alpha}, \ \ H\to \bar{\alpha}H,\ \ \phi\to\frac{1}{2\alpha_s^2}\phi,\ \ 
\phid\to\frac{1}{2\alpha_s^2}\phid,
\eeq
where standardly $\bar{\alpha}=\alpha_sN_c/\pi$.
In these variables, for $0<y<Y$,  the equations of motion have the same form
(9) and (10) without the coefficient before the nonlinear terms and with zero
right-hand side.  All parts of the action aquire a common coefficient $1/(2\alpha_s^2)$:
\beq
S_0=\frac{1}{2\alpha_s^2}
\langle\phid K\left(\frac{\partial}{\partial y}+H\right)\phi\rangle,
 \eeq
\beq
S_I=\frac{1}{2\alpha_s^2}\langle\Big({\phid}^2K\phi+\phi^2K\phid\Big)\rangle
\eeq
and
\beq
S_E=-\frac{1}{2\alpha_s^2}
\langle\phid K\phi \Big( \delta(y)+\delta(y-Y)\Big)\rangle,
\eeq
where we expressed the external sources via the boundary values of $\phi$ and $\phid$.
Note that the expression for $S_0$ assumes integration over all values of $y$, so that
the derivative in $y$ generates $\delta$-like terms which partially cancel with
the external part of the action. If one symmetrizes $S_0$ in $\phi$ and $\phid$ then
these terms cancel exactly one half of $S_E$. This implies that taking 
in $S_0$ the integration over $y$ in the interval $0<y<Y$ one has to take the total
action as
\beq
S=S_0+S_I+\frac{1}{2}S_E.
\eeq

The operator $K^{-1}$ appearing before the second non-linear term in (9) and (10) can be
represented as an integral operator in the transverse momentum space 
with a kernel ~\cite{bra1}
\beq
K^{-1}(q_1,q_2)=\frac{\pi}{2}\frac{1}{q_>^2}\left(\ln
\frac{q_>}{q_<}+1\right),
\eeq
where $q_{>(<)}=\max(\min)\{q_1,q_2\}$.
The operator $K$ contains the 4th derivative in q. To simplify it we present it
as a product 
\beq
K=L^{\dagger}L,\ \ L=q^2\nabla_q^2.
\eeq
In logarithmic variables $L$ reduces to the 2nd derivative. If 
\beq
q=q_0e^{\beta t}
\eeq
then
\beq
L=\frac{1}{\beta^2}\frac{\partial^2}{\partial t^2}.
\eeq
Using this  and integrating by part to exclude higher derivatives we can
express the complicated 2nd non-linear term in the equations via
finctions $\phi$ and $\phid$ and their first and second derivatives in $t$.
Denoting
\beq
\frac{\partial\phi}{dt}=\phi_1,\ \ \frac{\partial^2\phi}{\partial t^2}=\phi_2
\eeq
and similarly for $\phid$, we find 
\[
2K^{-1}\phid K\phi=
\frac{1}{2\beta^3}\Big\{\int_{-\infty}^{t}dt_1
e^{-2z}\phi_2\Big((z+1)\phid_2-2\beta\phid_1\Big)\]\beq
+\int_{t}^{\infty}dt_1\phi_2\Big((1-z)\phid_2+2\beta(2z-1)\phid_1
-4\beta^2z\phid\Big)\Big\},
\eeq
where $z=\beta(t-t_1)$.  The conjugated term has the same form
with $\phi\leftrightarrow\phid$.

Calculating the action one can split $K$ into a pair of operators $L$ acting on
factors depending on $\phi$ and $\phid$. In this way one obtains
\beq
S_0=\frac{1}{2\alpha_s^2\beta^4}<\phid_2\left(\frac{\partial}{\partial y}+H\right)\phi_2>
\eeq
(symmetrized in $\phi$ and $\phid$),
\beq
S_I=\frac{1}{2\alpha_s^2\beta^4}<2\phi_2(\phid_2\phi+\phi_1^2)+h.c>,
\eeq
\beq
S_E=-\frac{1}{2\alpha_s^2\beta^4}<\phid_2\phi_2\Big(\delta(y)
+\delta(y-Y)\Big)>.
\eeq

\subsection{Boundary conditions}
To fix our boundary conditions we use our experience with the non-linear BFKL
equation to study the nuclear structure functions ~\cite{bra2,arm}. The
adequate initial
values for $\phi(y,q)$ were taken there from the
Golec-Biernat-Wuesthoff distribution, fitted to the proton data at comparatively low
values of $x$ ~\cite{gbw}, which was duly eikonalized for a nucleus target. In fact
eikonalization implies including terms of  higher orders in $1/N_c^2$,
 outisde the precision of the approach. Also it is not clear how to generalize 
eikonalization procedure to the nucleus-nucleus case. For both of these
reasons our first choice (I) for the  initial values is the non-eikonalized Golec-Biernat-
Wuesthoff
distribution for the nucleus:
\beq
\phi(0,q)=-\frac{1}{2}a\,{\rm Ei}\,\left(-\frac{q^2}{0.21814}\right).
\eeq
Here $a$ carries information about the nucleus and impact parameter
\beq
a=\sigma_0T_A(b_A),
\eeq
$\sigma_0=20.8$ mb and $q$ is in GeV/c.
The value of $\phid(Y,q)$ was taken in the same form with $T_A(b_A)\to T_B(b_B)$.
To study a possible influence of the form of the initial distribution in $q$ we also
used an alternative choice  (II) with the same infrared behaviour and point
where the gluon distribution is peaked
but a  much slowlier fall of the distribution at large $q$
\beq
\phi(0,q)=-\frac{1}{2}a\ln\left(1+\frac{0.21814}{q^2}\right).
\eeq

\subsection{Iterative solution}
Our first method to find the stationary point of the  action has
been to solve
the classical equations of motion iteratively. We have chosen the sum of
pure fan diagrams as a starting function for iterations.
In practice this means that we first solve the equations with the non-linear term
mixing $\phi$ and $\phid$ put to zero. These solutions seve as an input
for the iterations $\phi^{(o)}$ and ${\phid}^{(0)}$.
Then we find next iterations from the equations
\beq
\left(\frac{\partial}{\partial y}+H\right)\phi^{(n+1)}+
{\phi^{(n+1)}}^2+2K^{-1}{\phid}^{(n)} K\phi^{(n)}=0
\eeq
and
\beq
\left(-\frac{\partial}{\partial y}+H\right){\phid}^{(n+1)}+
{{{\phid}^{(n+1}}}^2+2K^{-1}\phi^{(n)}  K{\phid}^{(n)}=0
\eeq
For each iteration we have only to evolve the initial function from $y=0$ to
$y=Y$, rather than solve the equivalent pair of two dimensional non-linear 
integral equations, which considerably diminishes computer time.

Unfortunately our calculations show that this method works only for  rather
small values of the participant atomic numbers  and rapidity $Y$.  Obviously the 
maximal value of the factor $a$  entering (29) or (30)
is achieved at $b=0$ and
for $\max\{A,B\}$. With the Woods-Saxon nuclear density for Pb-Pb, Cu-Cu,
Al-Al, O-O and C-C  collisions this $\max\{a\}$ is found to be equal
2.20, 1.53,1.11, 0.88  and 0.83 respectively.
Our calculations show that for $a\leq 2.2$ the described iteration procedure is convergent
only up to $Y=1.1$ for choice I of the initial distribution and up to $Y=1.8.$ for choice II.
For lighter nuclei, taking $a\leq 1.0$ we find that iterations converge up to $y=1.3$
for choice I and up to $Y=2$ for choice II.
The physical values of these rapidities depend on the chosen value of $\bar{\alpha}$.
With $\bar{\alpha}=0.2$ the above numbers are to be multiplied by factor 5, but still
remain rather low: for, say, O-O collisions iterations allow to move only
up to rapidities
of the order 10, that is c.m. energies of the order $150$ GeV and for Pb-Pb collision
the upper limit for the c.m. energy lowers to $\sim 90$ GeV.

\subsection{Variational solution}
A clear alternative is obviously to try to directly find the stationary point 
of the action
choosing some trial fields $\phi(y,q)$ and $\phid(y,q)$ which satisfy the
boundary conditions.
The difficulty of this approach is related to the fact that the action can have more
than one stationary point. 
In our first attempt we have chosen the simplest form for the trial fields with $y$ and $q$
dependence factorized. Moreover for the the $y$ dependence we chose a simple exponential one,
with a variable slope $\Delta$, so that our trial fields have the form
\beq
\phi(y,q)=e^{\Delta y}\phi(0,q),\ \ \phid(y,q)=e^{\Delta(Y-y)}\phid(Y,q)
\eeq
The boundary functions $\phi(0,q)$ and $\phid(Y,q)$ were taken according to 
(29) and (30) for variants I and II. 
The only variational parameter $\Delta$ was chosen to give the minimal 
value for action $S$.
Note, that with the fields having a simple analytic form, the necessary 
derivative functions
entering Eqs. (26)-(28) are easily found, so that calulating the action 
reduces to just
doing two independent integrations over $y$ and $q$.
With these trial fields  the solution for $\Delta$ always exists for any
values of $Y$ and
parameter $a$ in Eqs. (29)-(30) and moreover it corresponds to tne minimum 
of the action.
The quality of this approximation can be checked at $y$ and $a$ where the 
exact solution 
can be found perturbatively. At $Y=1$ and $a=1$ the exact values of action 
$S$ (without factor
$1/(2\alpha_s^2))$ are $-0.0120$ and $-0.0370$ for variants I and II
respectively.
The variational values obtained with (34) for these two variants are
$-0.0100$ and $-0.0262$.
As one observes the precision is not very high (especially for variant II).
Still we hope that
the variational approach might give some indication about the behaviour of 
the cross-section
and eikonal functions at large values of $Y$ at which we cannot obtain the 
exact solution.

\section{Numerical results}
We first report on the iterational solution of the field equations
(32),(33), which 
as mentioned is convergent at not too high values of $Y$, $A$ and $B$. We chose
to study O-O scattering ($A=B=16$) using choice II of the initial functions, 
which allowed us to obtain the solution up to $Y=2$. Presenting our results
we first consider
 the gluonic density, which can be related to functions $L\phi(y,q)=h(y,q)$
and $L\phid(y,q)=h^{\dagger}(y,q)$.
Indeed as follows from the study of the non-linear BFKL equation
the gluon density of a single heavy nucleus is given by ~\cite{bra2}
\beq
\frac{dxG(x,q)}{d^2bd^2q}=\frac{N_c}{2\pi^2\alpha_s}h(y,q),\ \ 
y=\bar{\alpha}\ln\frac{1}{x}
\eeq
We conjecture that the gluon density in the nucleus-nucleus collision
at rapidity $y$ will be given by a similar formula with contributions 
from both nuclei. For central collisions then
\beq
\frac{dxG(x,q)}{d^2bd^2q}=\frac{N_c}{2\pi^2\alpha_s}
(h(y,q)+h^{\dagger}(y,q)\Big), 
\eeq
Note that in the considered symmetric case  $\phid(y,q)=\phi(Y-y,q$
and so $h^{\dagger}(y,q)=h(Y-y,q)$.

In Fig. 2 we present our solution for $h(y,q)$ at $Y=2$, $b_A=b_B=0$
at  different stages of evolution:  $y=0$, 0.5, 1.0, 1.5 and 2.
For comparison we show in Fig. 3 the same function which is found from
the non-linear evolution equation for a single nucleus (only fan diagrams).
The difference between Fig. 2 and 3 comes from the influence of another
nucleus on the evolution process. As one can conclude, this
influence is quite strong. Whereas the fan-diagram density steadily
shifts towards higher momenta more or less preserving in  its shape,
the density for the nucleus-nucleus collision  practically does not move until
$y=1$, its peak dramatically falling and its low momentum tail visibly growing.
Only as late as at $y=1.5$ one notices some slow shift towards higher
momenta, which becomes more pronounced at $y=2$. Still at $y=2$ its peak
lies at $q=Q_s=2$ GeV/c (``saturation momentum''), whereas for single
nucleus it is found at  $Q_s=8$ GeV/c. In Fig. 4 we illustrate the
total gluon density in O-O collisions at $y=2$ given by the sum
$h(y,q)+h(Y-q)$ up to a factor depending on the coupling constant value.
This density has its peak at the point close to 1 GeV/c practically
independent of rapidity. The height of the peak is falling towards the
central region. Some diffusion towards small momenta is observed.
It can however hardly be compared to the diffusion for the pure BFKL evolution
illustrated in Fig. 5 for the same initial function and same region of $y$.

A clear physical observable is of course the total nucleus-nucleus
cross-section obtained by the integration of (1) over all impact
parameters.
We show it for O-O scattering at $Y\leq 2$ in Fig. 6. To compare we present
also the cross-sections corresponding to a single BFKL exchange.
These latter are naturally larger but the difference is not at all dramatic,
reaching some 18\%  at the maximal rapidity $Y=2$. This is understandable,
having in mind that Eq. (1) actualy automatically unitarizes the amplitude
and leads to very similar results even for very different eikonal
functions provided they are large. This can be clearly seen from the
comparison of the eikonal functions at $b=0$ in Fig. 7. At $Y=2$ the
pomeron interaction in the nucleus-nucleus collisions reduces it by
a order of magnitude, although it still remains large, $\sim 200$.
Some spreading of the gluon distribution into the low momenta domain
visible in Fig. 4 in the central rapidity region makes one think that the
results may be rather sensitive to the infrared region and so
strongly dependent on the infrared cutoff. Such a dependence indeed exists
but is not so strong. With the infrared cutoff at $k_{min}=0.3$ GeV
we obtain at $Y=2$ $\sigma_{O-O}=4.18$ bn and $T_{O-O}(0)=179$
whereas without cutoff we have $\sigma_{O-O}=5.36$ bn and
$T_{O-O}(0)=208$. As one observes the cross-section results more
sensitive to the infrared cutoff, which is a result of peripheral
collisions, where evolution follows the linear BFKL equation. 

Our variational results for both variants of the choice of the initial
functions are presented in Figs. 8-11. Since in this approximation we are 
not restricted to small values of $A$ and $B$, we show our results for 
Pb-Pb collisions ($A=B=207$) 
In Figs. 8 and 9 we show the total cross-sections for variants I and II
of the initial fields. They are again compared to the cross-sections
corresponding to the single BFKL exchange. All the cross-sections 
steadily rise with $Y$. However this rise seem to be very weak for the
choice I of the initial function. The single BFKL exchange naturally
leads to larger cross-sections and the ratio of these to the cross-sections
with pomeronic interaction rises with $Y$ reaching values 2 and 1.5
at $Y=6$
for variants I and II respectively.
However this difference is far larger for the eikonal functions,
shown in Figs. 10 and 11 at $b=0$. The eikonal function for
a single BFKL exchange rises up to values of the order 10$^9$ 
at $Y=6$, whereas with pomeronic interactions we find values 
around 100 or 1000 for variants I and II. It is remarkable that,
with the pomeronic interaction switched on, the eikonal function
actually diminishes 
with $y$ for central collisions. Therefore the rise of the cross-section is 
totally due to peripheral collisions, where, with a  low nuclear density,
the non-linear effects are small and the fields grow according to the
pure linear BFKL equation.      

\section{Discusssion} 
We have made the first attempt to solve the equations which describe 
nucleus-nucleus scattering in the framework of the perturbative QCD
with a large nucmber of colours and a fixed coupling constant.
The natural iterative approach has been found to converge in a 
restricted domain of not too high scaled rapidities and atomic
number of participants. Physical rapidities covered by the
convergence range depend on the value of the coupling constant.
For $\alpha_s=0.2$ they are not greater than 10 for
O-O collisons and not greater than 9 for Pb-Pb collisions.
The solutions in this range  of $Y$  has been found to generate
the gluon density which falls with $y$ (towards the central rapidity region)
and somewhat spreads into the infrared region of transverse momenta, 
its maximum staying around 1 GeV/c. It radically differs from the density
of the isolated nucleus, which is known to steadily shift towards higher
momenta, the hight of its  peak practically independent of $y$.
Both the eikonal function and the total cross-section are found to
be damped  as compared to the single pomeron exchange (by an order
of magnitude for the eikonal function). The latter has been found to
actually fall with energy for central collisions. However the total cross-sections
rise with energy due to peripheral collisions where non-linear effects are
naturally small.

Unfortunately we have not been able to find the solutions outside the mentioned
restricted domain of rapidities and atomic numbers. We do not know what sort
of singularity occurs at the boundaries of this domain and even if the solutions
of our equations exist at all. It is possible that a sort of phase transition occurs
at this boundaries, so that the equations have to be changed.

Just to see some qualitative features of a possible solution at high rapidities
and atomic numbers we applied a simple variational procedure, approximating
the fields by certain simple trial functions with a single parameter to be determined
from the stationary point equation. It is hopeless to expect to study the gluon
distribution from such a simple approach. One expects more reasonable answers
for the eikonal and especially for the total cross-sections which are weakly
dependent on moderate variations of the fields. Our results seem to indicate
that with the further rise of energy the cross-sections continue to grow slowly,
much slowlier than with a single BFKL exchange. The eikonal function in the center
continues to fall, very slowly in variant II for the initial function and rather fast
for variant I.

The main lesson to be learned from these first calculations is that the
dynamics of nucleus-nucleus collisions is much more complicated than for
collisions of a small probe on a single nucleus. The gluon densities
we have found have a much more complicated form than in the latter case
when they scale with the saturation momentum which grows with energy as a 
power. No scaling of this sort have been observed. The remaing problem is to
understand the reason of the breakdown of the iterative solution at a certain
value of energy and/or atomic number and to try to move beyond this value.

\section{Acknowledgments}.
The author is most thankful to B.Vlahovic for his  constant interest in
this work and discussions.. He is also thankful to the North Carolina Central 
University, USA, for financial support.

\section{Figure captions}

Fig. 1. Some pomeron tree diagrams summed by our effective field theory.

Fig. 2. The gluon distribution corresponding to the field $\phi(y,q)$ with 
initial function II. Numbers show rapidities $y=0$, 0.5, 1., 1.5 and 2.

Fig. 3. Same as Fig. 2 for the field evolving according to
the non-linear BFKL equation (only fan diagrams). Curves from left to right 
correspond to $y=0$, 0.5, 1. 1.5 and 2.

Fig. 4. The total gluon density in O-O collisions. Numbers show rapidity 
distances from the target or projectile: $y=0$, 0.5 and 1.

Fig. 5. Same as Fig. 2 for the field evolving according to the
linear BFKL equation. Curves from bottom to top correspond to rapidities
$y=0$, 0.5, 1., 1.5 and 2.

Fig. 6. The total cross-section for O-O collisions with initial 
conditions II (lower curve). The upper curve corresponds to a single 
BFKL exchange.

Fig. 7. The eikonal function at $b=0$ for O-O collisions with
initial conditions II (lower curve). The upper curve corresponds to a
single BFKL exchange.

Fig. 8. Variational estimates for the total cross-sections for Pb-Pb 
collisions at high rapidities with initial conditions I (lower curve).
The upper curve corresponds to a single BFKL exchange.

Fig. 9. Same as in Fig. 9 for initial conditions II.

Fig. 10. Variational estimates for the eikonal function at $b=0$  
for Pb-Pb 
collisions at high rapidities with initial conditions I (lower curve).
The upper curve corresponds to a single BFKL exchange.

Fig. 11. Same as in Fig. 10 for initial conditions II.
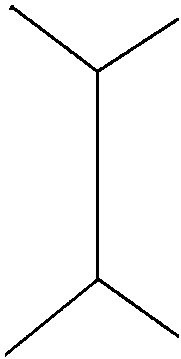
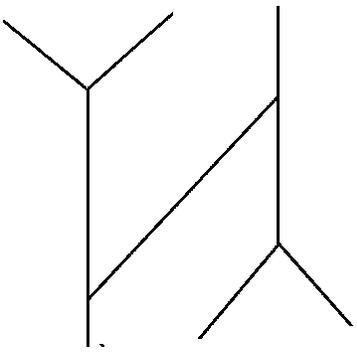
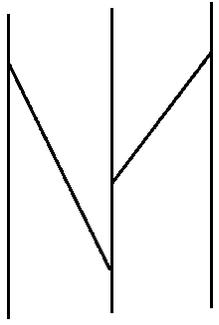
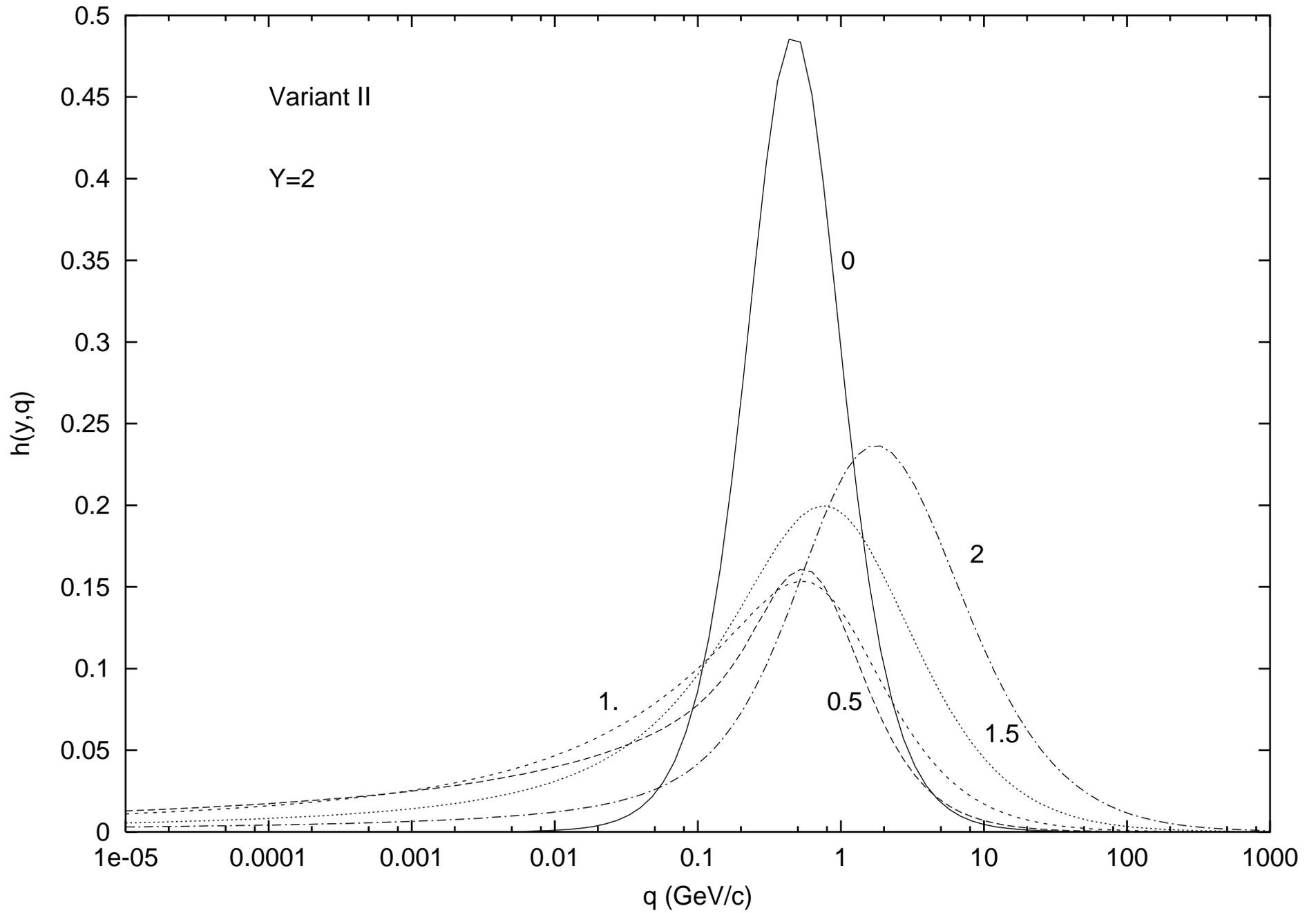
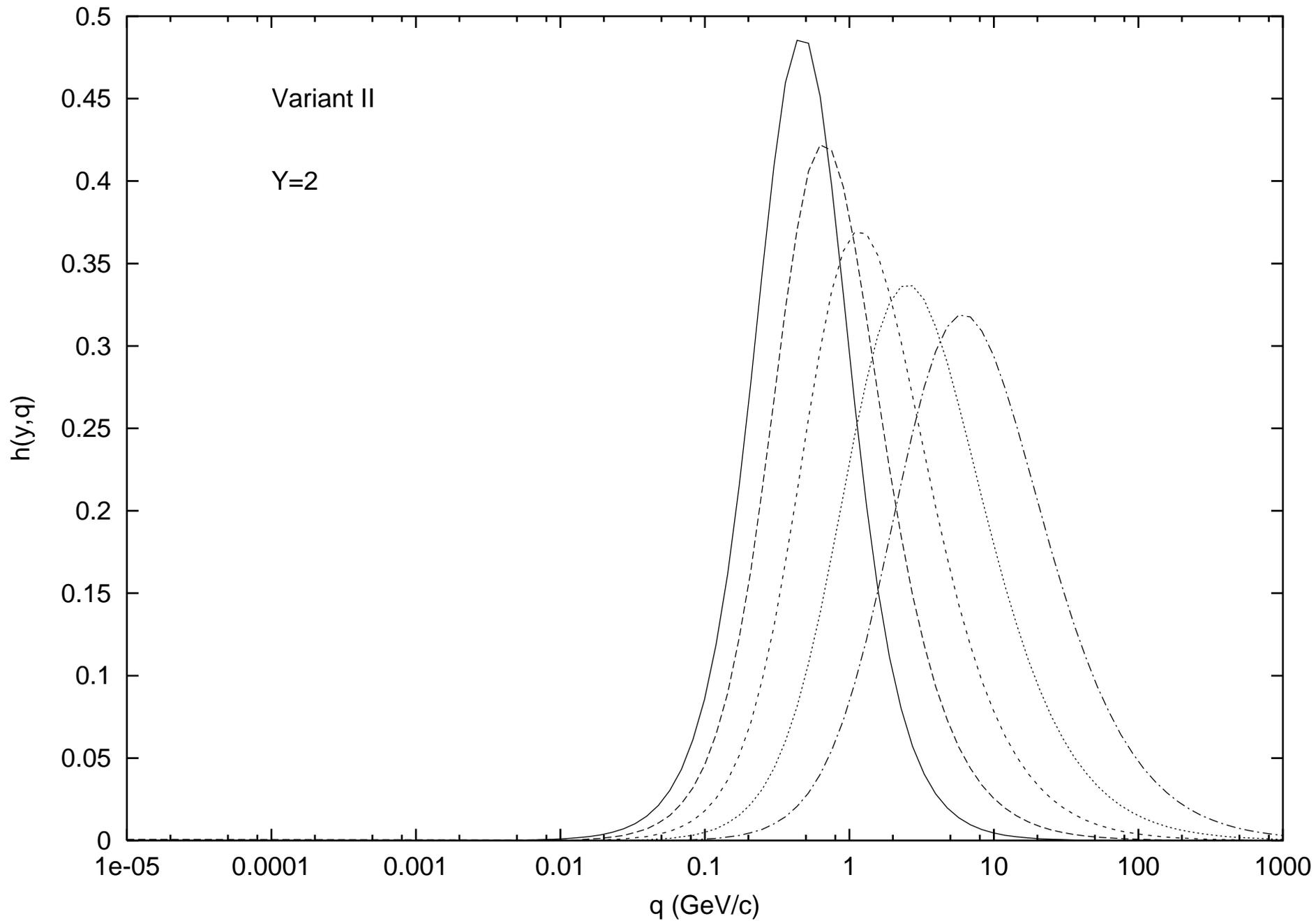
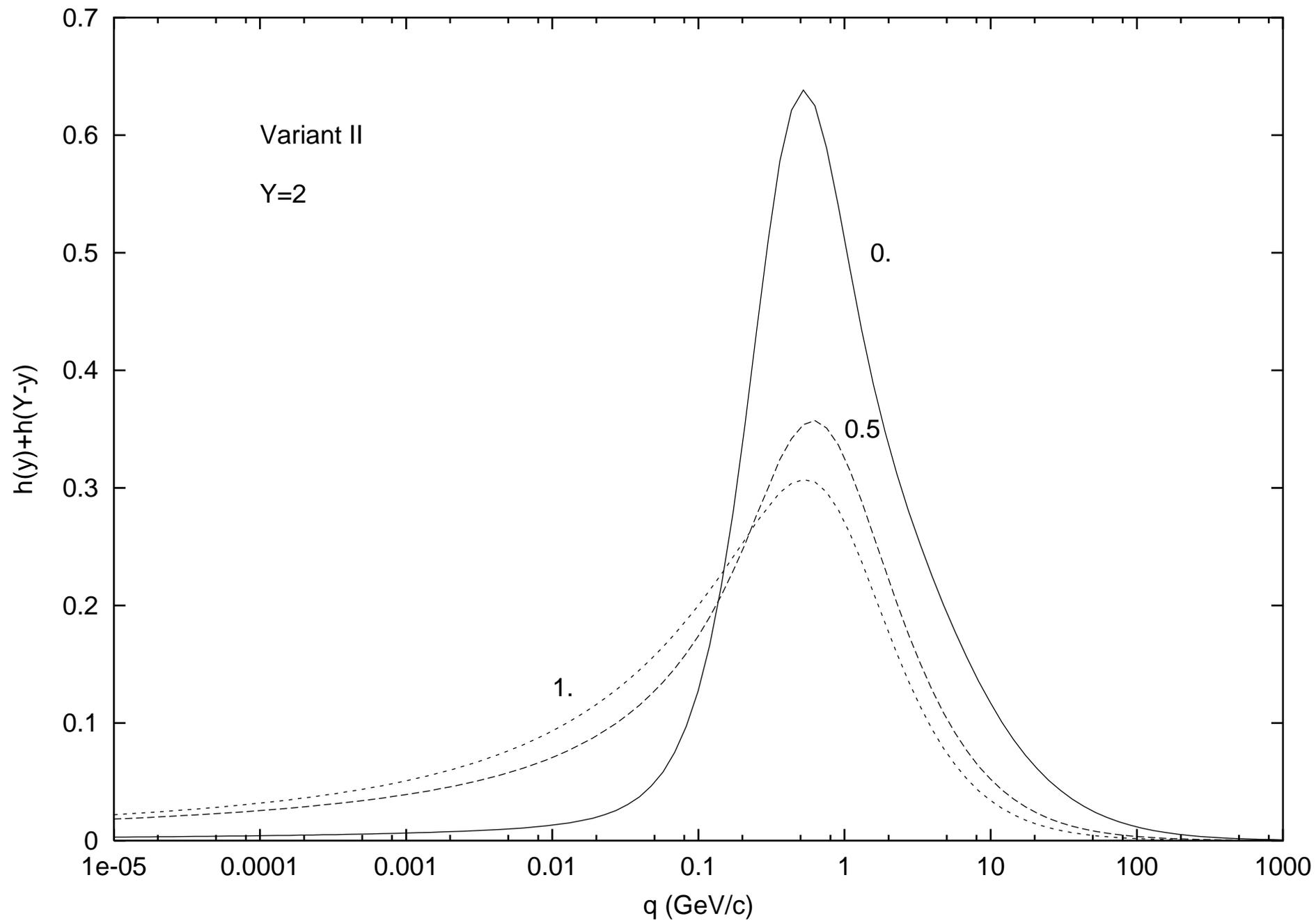
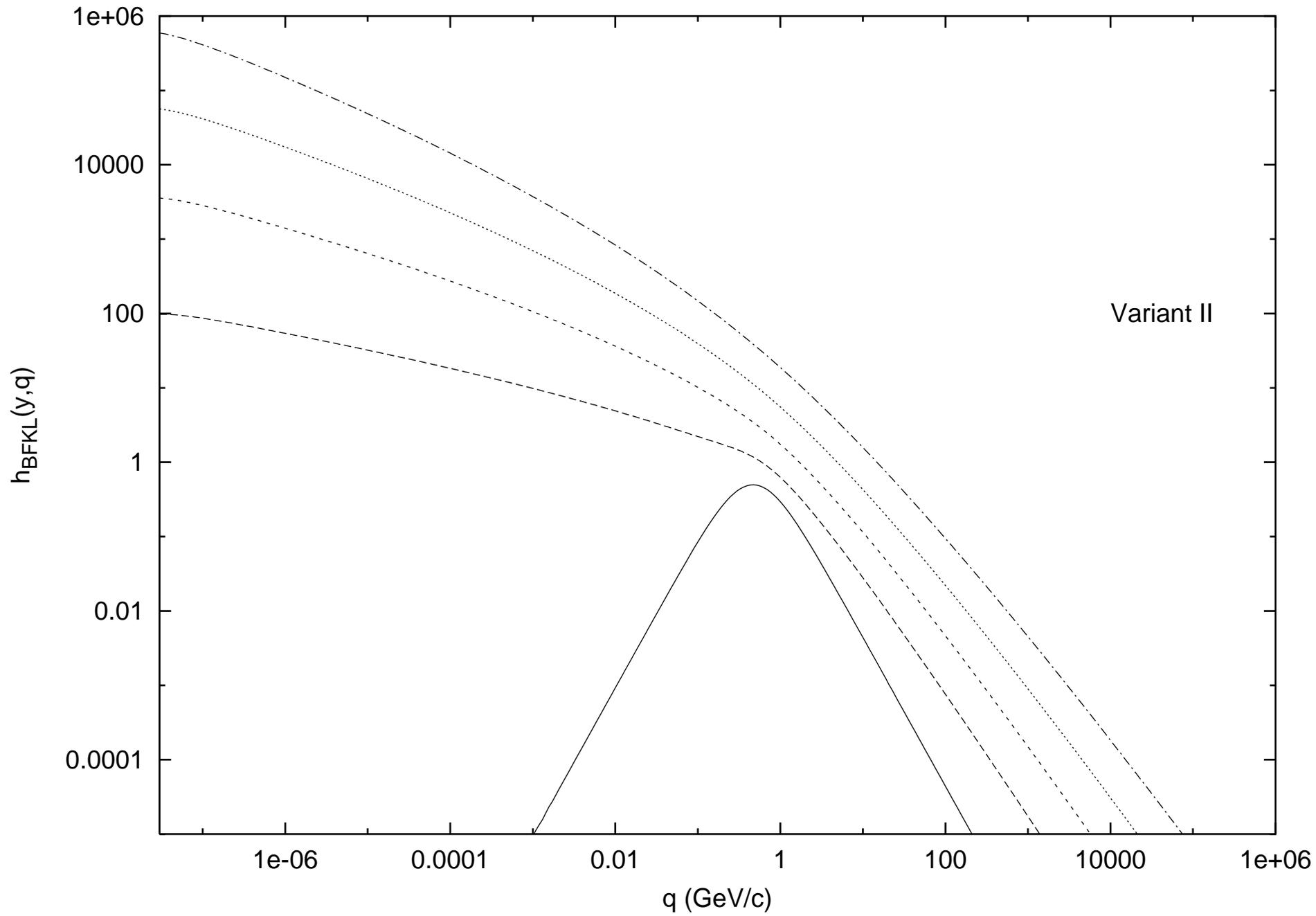
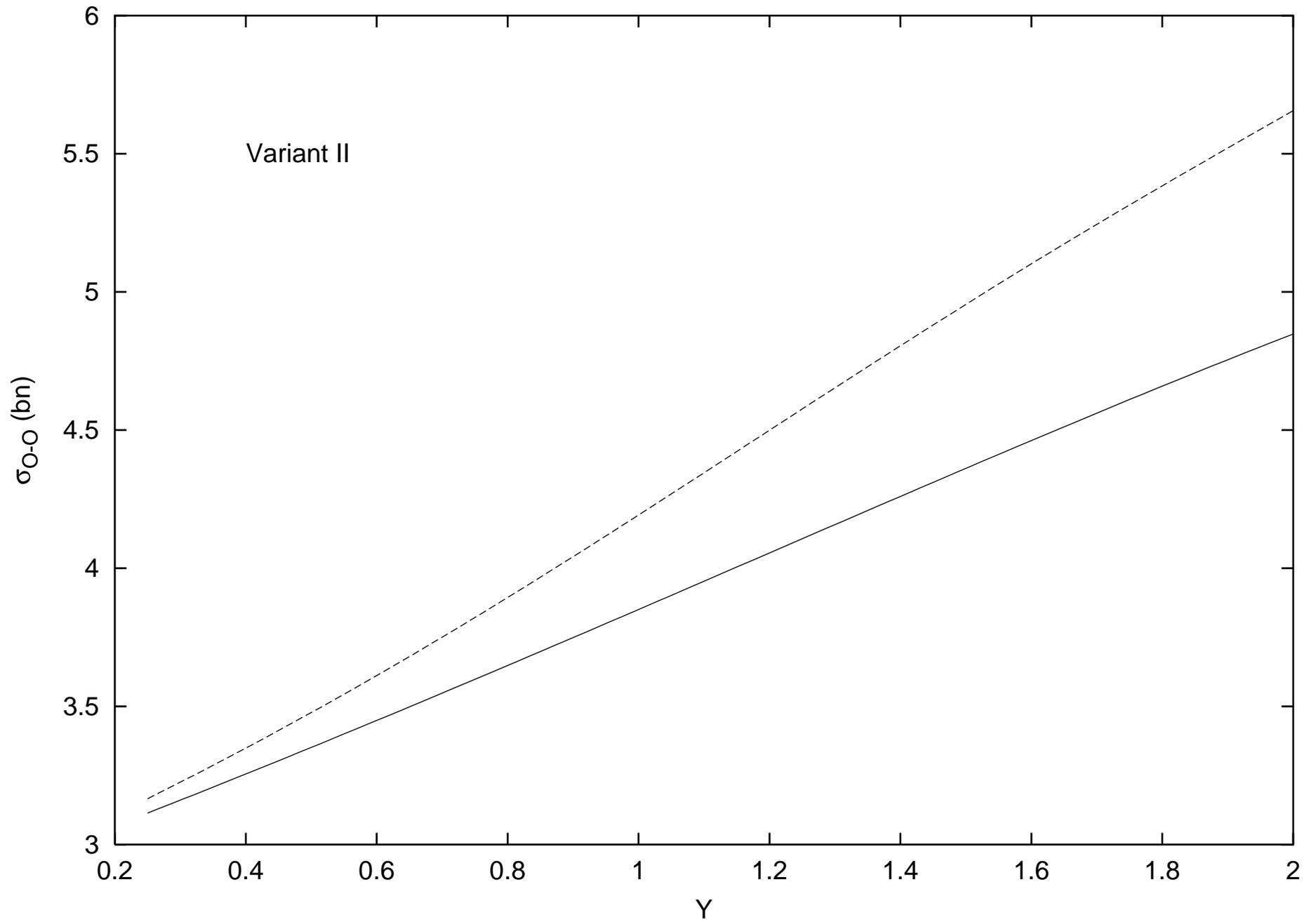
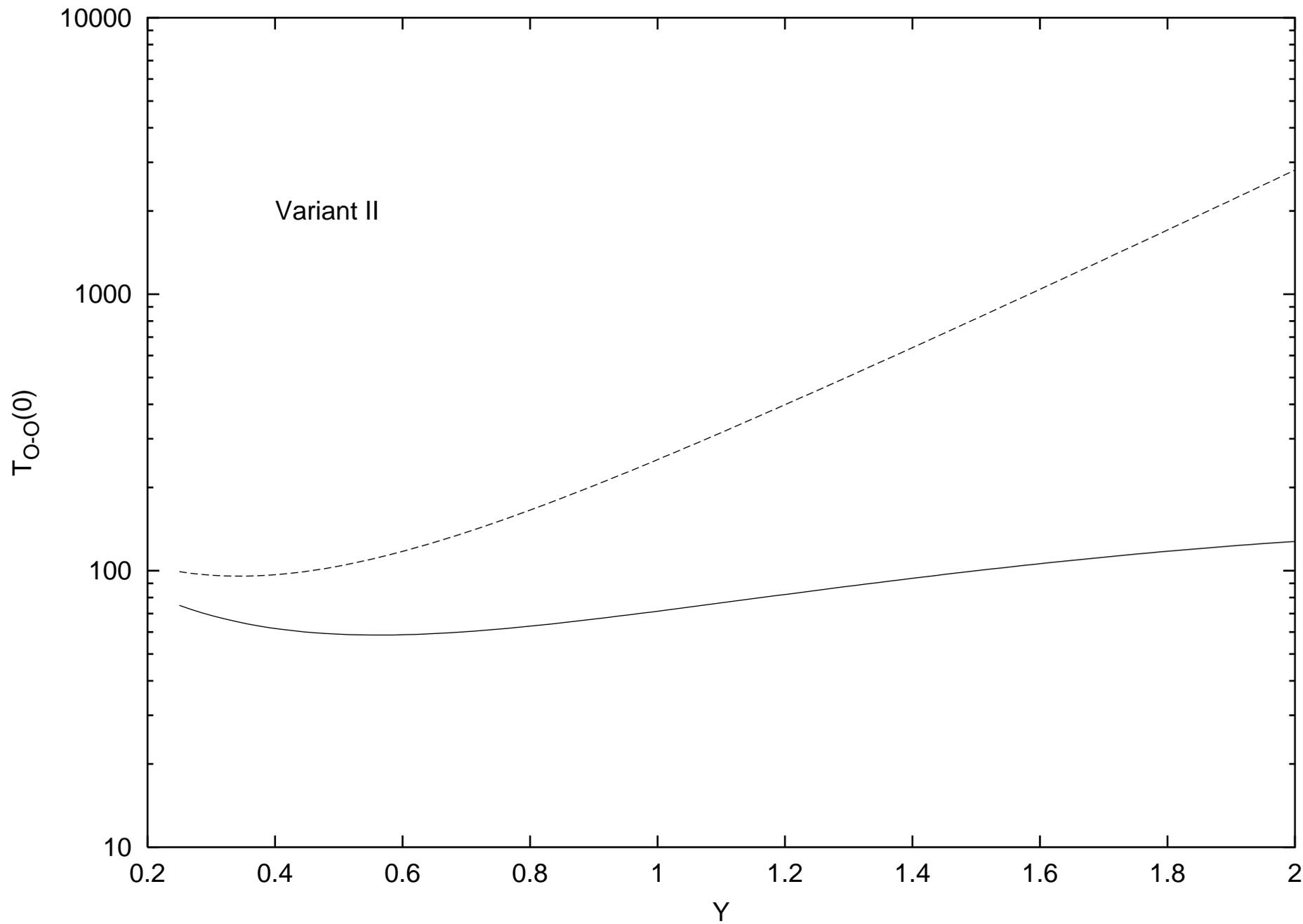
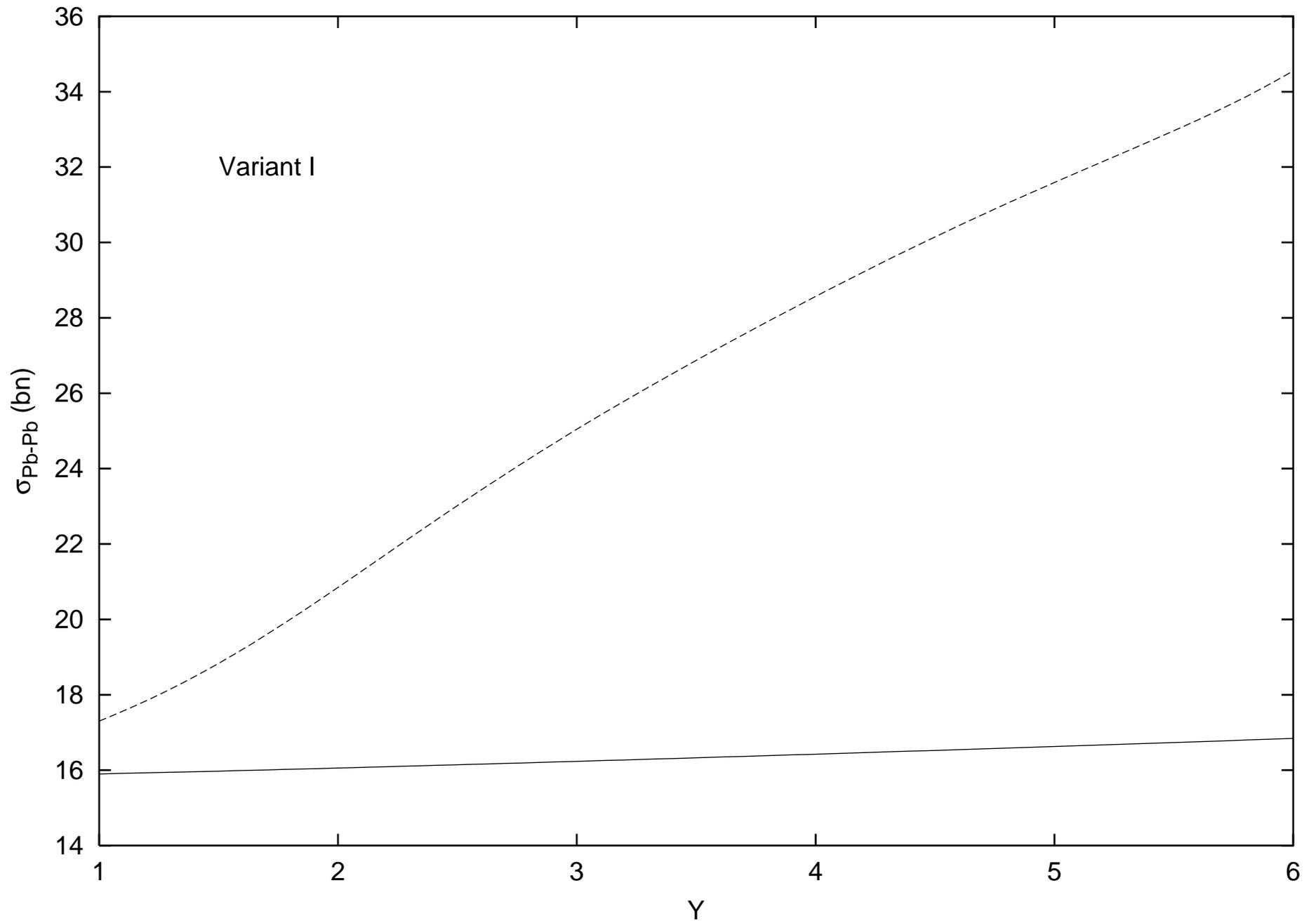
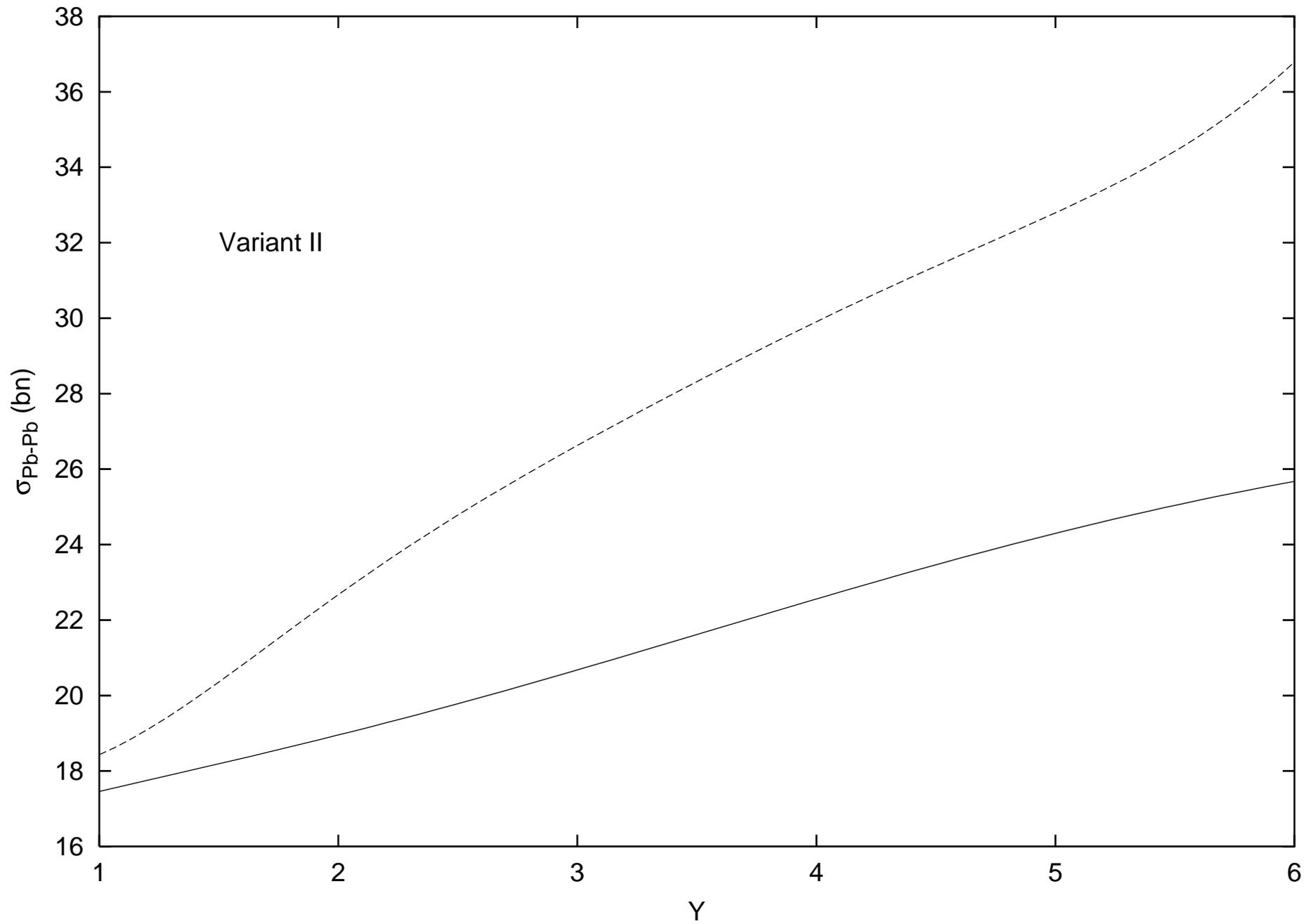
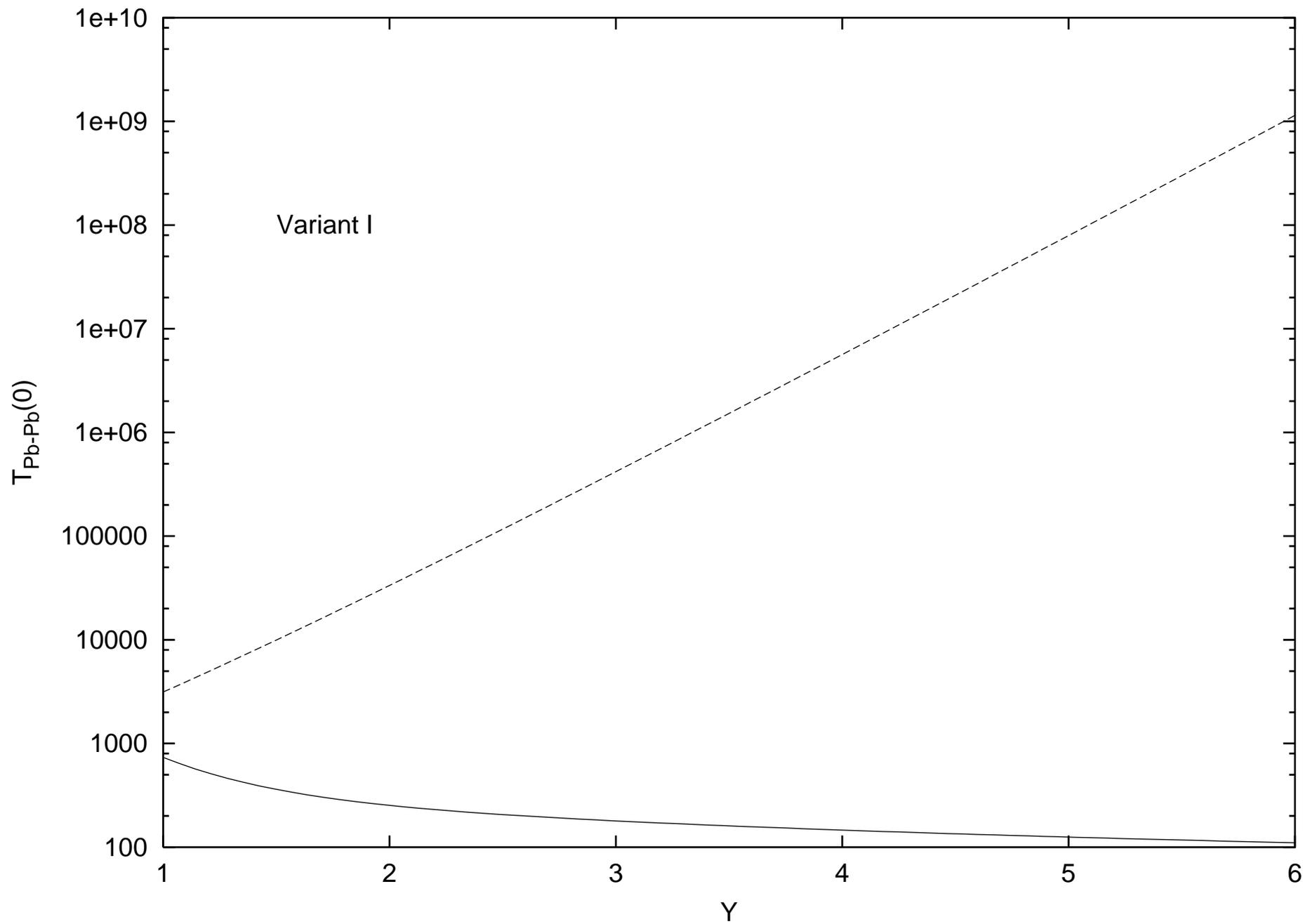
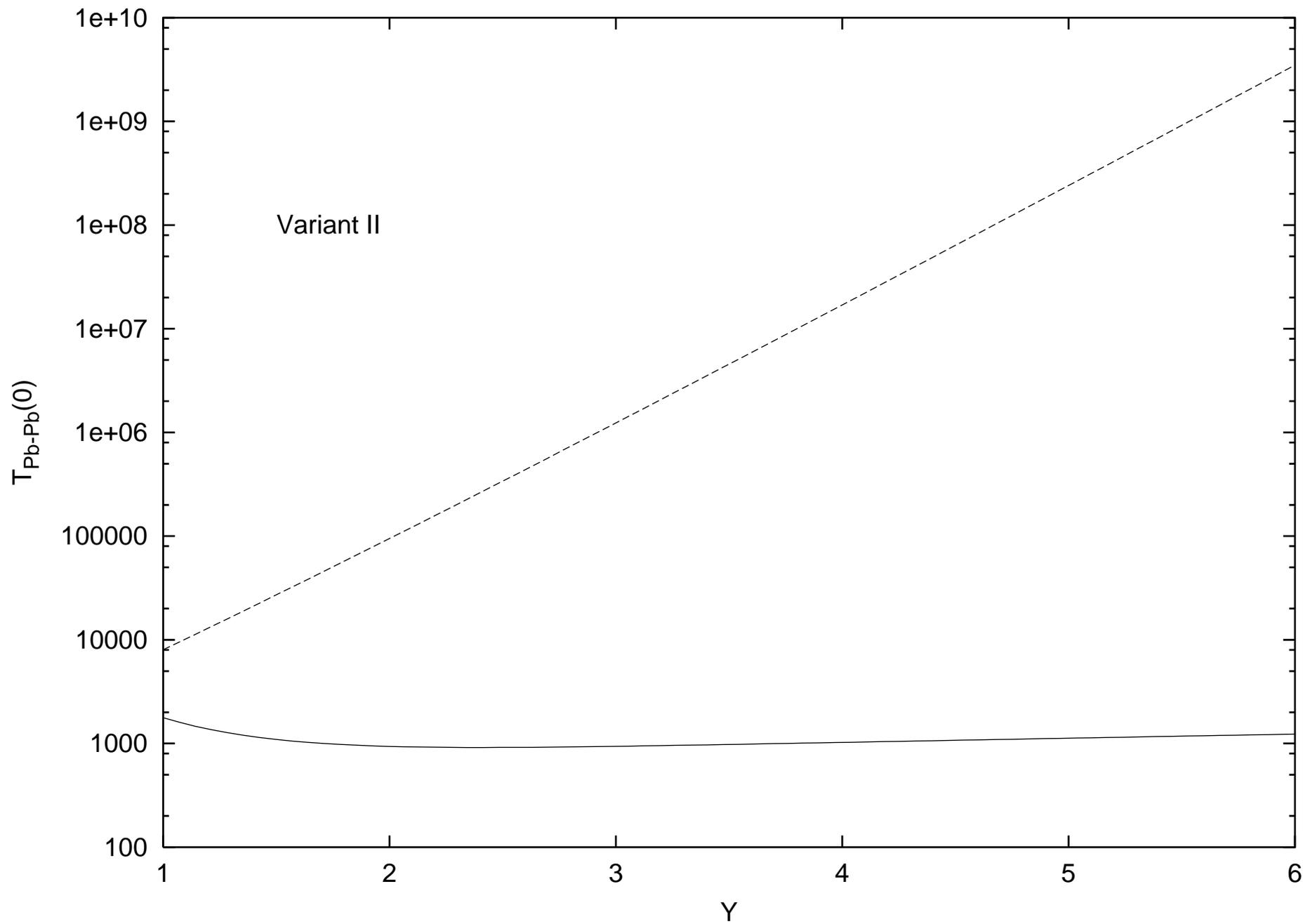
\end{document}